\documentclass[journal=nalefd,manuscript=article,layout=twocolumn]{achemso}
\usepackage{chemformula} % Formula subscripts using \ch{}
\usepackage[T1]{fontenc} % Use modern font encodings
\usepackage[colorlinks,linkcolor=blue,citecolor=blue,urlcolor=blue]{hyperref}
\usepackage{amsmath,mleftright,mathtools,bm,amssymb}

\def\a{\alpha}
\def\b{\beta}

\def\h{\eta}

\def\w{\omega}
\newcommand{\hphi}{\hat{\phi}}
\def\G{\Gamma}
\def\callG{\mbox{$\mathcal{G}$}}

\def\dd{\mathrm{d}}
\def\Tr{{\rm Tr}}
\def\Re{{\rm Re}}

\def\bgm{\bm\mu}
\def\bgn{\bm\nu}

\author{R. Tuovinen}
\affiliation{Department of Physics, Nanoscience Center,\\ P.O. Box 35, 40014 University of Jyv{\"a}skyl{\"a}, Finland}
\email{riku.m.s.tuovinen@jyu.fi}

\author{Y. Pavlyukh}
\affiliation{Institute of Theoretical Physics, Faculty of Fundamental Problems of Technology,\\
Wroclaw University of Science and Technology, 50-370 Wroclaw, Poland}
\email{yaroslav.pavlyukh@pwr.edu.pl}

\title{Electroluminescence rectification and high harmonic generation in molecular junctions}


\begin{document}
\begin{abstract}
The field of molecular electronics has emerged from efforts to understand electron propagation through single molecules and to use them in electronic circuits. Serving as a testbed for advanced theoretical methods, it reveals a significant discrepancy between the operational time scales of experiments (static to GHz frequencies) and theoretical models (femtoseconds). Utilizing a recently developed time-linear nonequilibrium Green's functions formalism, we model molecular junctions on experimentally accessible timescales. Our study focuses on the quantum pump effect in a Benzenedithiol molecule connected to two copper electrodes and coupled with cavity photons. By calculating both electric and photonic current responses to an ac bias voltage, we observe pronounced electroluminescence and high harmonic generation in this setup. The mechanism of the latter effect is more analogous to that from solids than from isolated molecules, with even harmonics being suppressed or enhanced depending on the symmetry of the driving field.
\end{abstract}

\paragraph{Introduction.}
Molecular systems are prospective elements of future electronic devices~\cite{flood_whence_2004,chen_from_2021}. One of the most interesting functionalities is their role as nano-junctions.~\cite{solomon_exploring_2010, lissau_tracking_2015, miwa_many-body_2019, wang_plasmonic_2022} State-of-the-art calculations are able to accurately predict ground- and excited-state properties of technologically relevant molecules. However, \emph{ab initio} description of photo-assisted tunneling, optical rectification, and electrically driven photon emission requires a new set of tools.  

The theory of quantum transport was initially developed by Landauer and B{\"u}ttiker,~\cite{landauer_spatial_1957, buttiker_generalized_1985, buttiker_four-terminal_1986} and further extended by Meir, Wingreen, and Jauho~\cite{meir_landauer_1992, jauho_time-dependent_1994}. They introduced a formula to calculate electric current in correlated junctions using nonequilibrium Green's functions (NEGF), which is applicable both in transient and stationary regimes. However, applying the Jauho-Meir-Wingreen formula is computationally challenging due to NEGF's intricate two-time structure, which hinders real-time simulations~\cite{cohen_greens_2020, tuovinen_electronic_2021, ridley_many-body_2022}.

Recently, we introduced a time-linear scaling Green's function approach for simulating open and correlated quantum systems out of equilibrium~\cite{tuovinen_time-linear_2023}. This method, rigorously tested on correlated lattice models~\cite{pavlyukh_cheers_2023}, has proven highly efficient and applicable to realistic 3D systems like molecules. In comparison with density functional theory (DFT)-based methods~\cite{lang_resistance_1995, kurth_transport_2017}, our approach offers distinct advantages by accommodating collision terms for correlated electrons and bosons. Furthermore, it differs technically from transport calculations reliant on time-dependent DFT, where Kohn-Sham orbitals are evolved in time~\cite{kurth_time-dependent_2005}.

\begin{figure}[t]
 \includegraphics[width=\columnwidth]{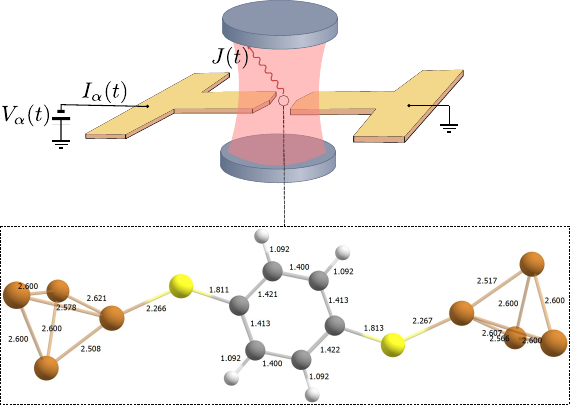}
  \caption{\small Setup and molecular geometry used in the calculations. A BDT molecule (finite quantum system) is contacted to macroscopic copper electrodes, and the electrons within the molecule are coupled with cavity photon fields. The molecular junction is driven out of equilibrium by time-dependent voltages $V_\a(t)$ on the electrodes $\alpha$, which generate time-dependent charge current $I_\a(t)$ and photon energy flux $J(t)$. Structural parameters in {\AA} obtained from all-electron DFT geometry optimization performed with the \textsc{orca} program~\cite{neese_orca_2012} using Perdew-Burke-Erzerhoff (PBE) GGA density functional~\cite{perdew_generalized_1996} and def2-TZVP~\cite{weigend_balanced_2005} basis set with six frozen terminal Cu atoms. Subsequent transport calculations are performed with \textsc{cheers} nonequilibrium Green's function code~\cite{pavlyukh_cheers_2023}.}
  \label{fig:geometry}
\end{figure}

Here, using the method from Ref.~\citenum{tuovinen_time-linear_2023}, we study electron current and photonic flux in a nano-junction of the $1{,}4$-Benzenedithiol (BDT) molecule connected to two copper (Cu) electrodes (Fig.~\ref{fig:geometry}). We observe significant current rectification, electrically-driven photon emission (the inverse of optical rectification~\cite{ward_optical_2010}), quantum pumping~\cite{arrachea_green-function_2005}, and strong upconversion, where photon energies exceed the applied field bias. Similar to how the efficiency of the famous Archimedes' screw depends on the tilting angle and the spiral step, the efficiency of the quantum pump depends on the magnitude and phase difference of the bias voltages, as previously demonstrated for model systems~\cite{ridley_time-dependent_2017,tuovinen_time-resolved_2019}. Additionally, we uncover effects related to the interaction with cavity photons. Given the flexibility to tune the strength of electron-photon coupling in optical cavities by changing the quantization volume~\cite{riso_molecular_2022}, we focus on the moderate coupling regime. For stronger couplings, correlated methods re-summing infinite number of diagrams, such as the $G\widetilde{W}$ method~\cite{pavlyukh_interacting_2022}, or even more expensive, numerically exact quantum Monte Carlo methods~\cite{cohen_taming_2015, erpenbeck_shaping_2023} may be necessary.

The BDT molecule has been extensively studied since the conductivity measurement of~\citet{reed_conductance_1997} using a mechanically controlled break-junction. Current--voltage, differential conductance, and inelastic electron tunneling spectroscopy measurements show that BDT is one of the very rare organic molecules able to span low to high transmission ranges, making it highly useful for molecular electronics applications and as a testbed for quantum transport~\cite{kim_benzenedithiol_2011}. It was used to test quantum chemistry calculations for molecules coupled to reservoirs~\cite{arnold_quantum_2007}, quantify the applicability of the wide-band limit in DFT transport calculations~\cite{verzijl_applicability_2013}, and to assess structural properties' influence on electronic transport with Au, Ag, and Cu electrodes~\cite{silva_structural_2020}. It was shown theoretically~\cite{bratkovsky_effects_2003} and then confirmed experimentally~\cite{kim_benzenedithiol_2011}, that current varies significantly with molecule orientation. \citet{matsuhita_conductance_2013} reported on vibrational modes and their shifts due to electrode interaction.

\paragraph{Model and method.}\label{sec:model}

We focus on nonequilibrium transport phenomena triggered by periodic zero net bias $V_\a(t)$ and observed through electronic currents $I_\a(t)$ and photonic energy flux $J(t)$~\cite{wang_plasmonic_2022} (Fig.~\ref{fig:geometry}). Our approach includes: i) a fully \emph{ab initio} treatment of the ground state of molecules coupled to electrodes, and ii) a time-linear scaling dynamics for open systems coupled with photons, using the wide band limit approximation (WBLA). While Floquet formalism has been used for studying long-term dynamics of periodically driven open systems~\cite{arrachea_relation_2006,stefanucci_time-dependent_2008}, our time-resolved approach also accommodates nonperiodic driving (e.g., short pulse excitations) and allows for the adiabatic switching procedure to prepare the initial correlated state~\cite{tuovinen_electronic_2021}.

In our calculations, we assume that the molecule interacts with three photonic modes with energies $\omega_{\bgm}$ coupled to the electron system via three-projections of the dipole moment operator, $g_{\mu, ij}= \sqrt{\omega_{\bgm}} d^{\bgm}_{ij}$.  Here, we follow the notation introduced in Ref.~\citenum{pavlyukh_time-linear_2022-1}, i.e., the greek index $\mu=(\bgm,\xi)$ specifies the bosonic mode and the component of the mode displacement vector: $\hphi_{\mu}=\hat{x}_{\bgm}$ for $\xi=1$ and $\hphi_{\mu}=\hat{p}_{\bgm}$ for $\xi=2$ with commutation rules $[\hphi_{\mu}, \hphi_{\mu'}]=\sigma_{\mu\mu'}$ and $\sigma_{\mu\mu'}=-\delta_{\bgm\bgm'} \left(\begin{smallmatrix}0  & -i \\i & 0 \end{smallmatrix}\right)_{\xi\xi'}$. 

Next, we consider a typical quantum transport setup, the finite system ($A$) being the junction and the electronic reservoirs being the electrodes (indexed by $\alpha=\{L,R\}$ and collectively denoted as $B$; with applied voltages $V_\a(t)$). The correlation effects in the electrodes and between the electrodes and the system are neglected, leading to the equation of motion (EOM) for $\rho_{ij}(t)=\langle \hat{d}_j^\dagger(t) \hat{d}_i(t)\rangle$\,---\,the electron density matrix~\cite{tuovinen_time-linear_2023}
\begin{align}\label{eq:rho:eom}
i\frac{\dd}{\dd t}\rho(t) & = h^{e}_{\rm eff}(t)\rho(t)+\frac{i}{4}\G(t) \nonumber \\
& + i\sum_{\ell\a} s_{\a}(t)\frac{\eta_\ell}{\beta}\G_{\a}\callG^{\rm em}_{\ell\a}(t) - {\rm h.c.},
\end{align}
where $h^{e}_{\rm eff}(t)\equiv h^{e}(t)-i\G(t)/2$ is the effective (non-self-adjoint) mean-field Hamiltonian, $s_{\a}(t)$ is a ramp function for the contact between the system and electrode $\a$, $\G_{\a}$ is the respective quasi-particle line-width matrix, and $\G(t)=\sum_\a s_\a^2(t)\G_\a$. The mean-field Hamiltonian $h^{e}(t)$ comprises the induced potentials due to the electron density and the cavity photons
\begin{align}\label{eq:hele}
   \!\! h^{e}_{ij}(t)&=h_{ij}+V^{{\rm HF}}_{ij}(t)+\sum_{\mu}g_{\mu,ij}s_\w(t)\phi_{\mu}(t),
\end{align}
where $h_{ij}$ is the single-particle part, and $V^{{\rm HF}}_{ij}(t)=\sum_{mn}[v_{imnj}-v_{imjn}]\rho_{nm}(t)$. $v_{imnj}$ are the Coulomb matrix elements and $g_{\mu,ij}$ are the electron-photon coupling coefficients. The ramp function $s_\w(t)$ is used in the adiabatic switching protocol to generate a correlated electron-photon initial state. Eqs.~\eqref{eq:rho:eom} and~\eqref{eq:hele} apply to the $A$ subsystem, omitting the $AA$ block designation for brevity. The pure electronic Hartree-Fock (HF) Hamiltonian is $h^\text{HF}_{ij}(t)=h_{ij}+V^{{\rm HF}}_{ij}(t)$. Photonic displacements $\phi_{\mu}$ are co-evolved according to the Ehrenfest equations of motion
\begin{equation}\label{eq:phi:eom}
i\frac{\dd}{\dd t}\phi_\mu(t) - \sum_{\nu} h^{\mathrm{ph}}_{\mu\nu}\phi_{\nu}(t) = \sum_{ij}\bar{g}_{\mu,ij}s_\w(t) \rho_{ji}(t),
\end{equation}
where we introduced the photonic Hamiltonian $h^{\mathrm{ph}}_{\mu\nu}=\sigma_{\mu\nu}\omega_{\bgn}$, and $\bar{g}_{\mu,ij}=\sum_{\nu}\sigma_{\mu\nu}g_{\nu,ij}$. 
Finally, $\callG^{\rm em}_{\ell\a}(t)$ is the embedding correlator propagated according to
\begin{align}\label{eq:gla:eom}
i\frac{\dd}{\dd t} \callG^{\rm em}_{\ell\a}(t) = - s_{\a}(t) & -\mathcal{G}^{\rm em}_{\ell\a}(t)\Big(h^{e\dag}_{\rm eff}(t) \nonumber \\
& -V_\a(t)-\mu+i\frac{\zeta_\ell}{\beta}\Big)
\end{align}
with $\mu$ being the chemical potential. The residues $\h_{\ell}$ and poles $\zeta_{\ell}$ result from an efficient expansion~\cite{hu_communication:_2010} of the Fermi distribution $f(\w)\equiv\frac1{e^{\beta \w}+1}= \frac{1}{2}- \sum_{\ell} \eta_\ell \big(\frac{1}{\b \w+i\zeta_\ell}+\frac{1}{\b \w -i\zeta_\ell}\big)$ with $\beta$ being the inverse temperature.

The main equations of our approach include the EOMs for the electron density matrix~\eqref{eq:rho:eom}, photonic displacements~\eqref{eq:phi:eom}, and the embedding correlator~\eqref{eq:gla:eom}, along with the mean-field electronic Hamiltonian~\eqref{eq:hele}. Derived within the nonequilibrium Green's function formalism using the generalized Kadanoff-Baym Ansatz (GKBA) and WBLA, these equations respect fundamental conservation laws, avoid time-convolutions, thus ensuring numerical evolution scales linearly with physical simulation time. 
The approximate GKBA+WBLA scheme is known to provide an accurate description of steady-state and transient currents when the electrochemical potential is well inside the electrode bandwidth~\cite{latini_charge_2014, tuovinen_electronic_2021, ridley_many-body_2022}.
Unlike more general theories~\cite{schluenzen_achieving_2020, karlsson_fast_2021, perfetto_real-time_2022, pavlyukh_interacting_2022, joost_dynamically_2022}, our formulation excludes computationally demanding electronic and photonic correlations, which would be necessary for strongly correlated systems.

Equations~\eqref{eq:rho:eom}-\eqref{eq:gla:eom} can, in principle, be solved in any basis, but a set of HF molecular orbitals is clearly of advantage since i) their compatibility with many-body perturbation theory, and ii) initially (when $A$ and $B$ subsystems are decoupled, i.e., $s_\a(t_i)=0$) the electron density matrix is diagonal and stationary. Our initial conditions are thus
\begin{align}
  \rho_{ij}(t_i)&=f_i\delta_{ij},&
  \phi_\mu(t_i)&=0,&
  \callG^{\rm em}_{\ell\a}(t_i)&=0,
\end{align}
where $f_i$ is the occupation of $i$th molecular orbital (the $i$ index comprises a spatial and a spin parts, i.\,e. $i\equiv(\bm{i},\sigma)$; $\bm{i}=1,\ldots,14$ with $7$ initially occupied states). Still, choosing the basis is nontrivial. One could argue that Kohn-Sham orbitals from frozen-density embedding~\cite{wesolowski_frozen-density_2015} or other quantum embedding theories~\cite{sun_quantum_2016} might better describe the correlated initial state and simplify partitioning between subsystems $A$ and $B$. Partition density functional theory, while formally exact for calculating molecular properties from the Kohn-Sham description on isolated fragments~\cite{cohen_foundations_2007}, requires fractionally occupied fragments~\cite{elliott_partition_2010}. In contrast, we prefer such charge transfer to arise naturally from transport theory, while also capturing long-range Coulomb interactions such as mutual polarization of the fragments.

\begin{figure}[t]
 \includegraphics[width=\columnwidth]{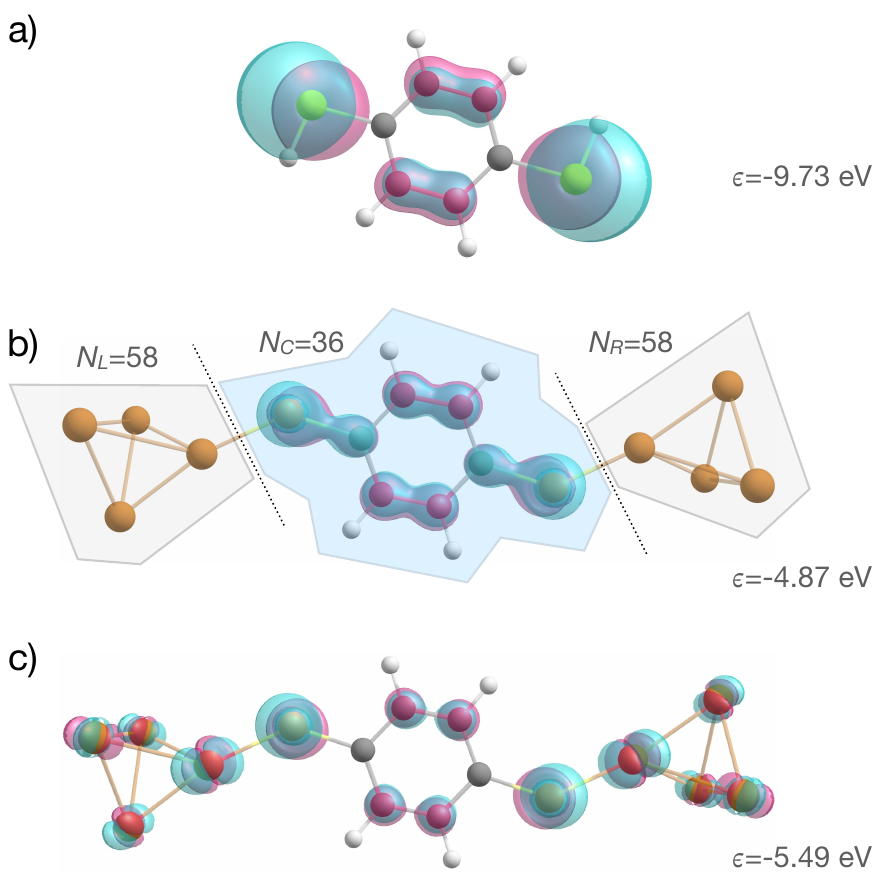}
  \caption{\small A representative molecular orbital included in our transport simulations, along with its energy as obtained from: a) stand-alone calculation of the BDT molecule; b) fragment calculation neglecting off-diagonal density blocks; c) full calculation of the \ch{Cu4-BDT-Cu4} molecule. Density localization is performed constraining the number of electrons in each fragment as indicated in b).}
  \label{fig:homo}
\end{figure}

To this end, we introduce coupled self-consistent field equations, in which the mean-field Hamiltonian $h^\text{HF}$ is constructed from the \emph{block-diagonal} density matrix
\begin{equation}
\rho = \begin{pmatrix}\rho_{AA} & 0 \\ 0 & \rho_{BB}\end{pmatrix}.
\end{equation}
The density matrix blocks are obtained by diagonalizing $h^\text{HF}_{AA}[\rho]$ and $h^\text{HF}_{BB}[\rho]$. The HF Hamiltonian for the central part ($A$) includes the kinetic part, interaction with nuclei of $A$ and $B$, and is partially neutralized by the electrostatic and exchange potential associated with $\rho_{AA}$ and $\rho_{BB}$. The same applies to the electrode part ($B$). Constructed via the Aufbau principle, $\rho_{AA}$ and $\rho_{BB}$ ensure no charge flows between fragments but polarization effects are included (Fig.~\ref{fig:homo}). The converged density matrix $\rho_0$ is obtained this way: The off-diagonal Hamiltonian blocks, $T\equiv h^\text{HF}_{AB}[\rho_0]$, determine the $\Gamma_\a$ matrix for electrode $\a$ as
\begin{align}
  \Gamma_{\a,ij} &= \sum_{a} \frac{\gamma}{\pi}\frac{T_{i,\a a}T_{j,\a a}}{(\epsilon_{\a a}-\mu)^2+\gamma^2},
\end{align}
where $\epsilon_{\a a}$ are the eigenvalues of $h^\text{HF}_{BB}[\rho_0]$, $\gamma$ is a small positive parameter, and $\mu$ is the chemical potential of $A$.

\paragraph{Initial-state preparation.}
In order to study time-resolved, correlated quantum dynamics in the molecular junction we prepare a coupled correlated initial state by adiabatically switching on the contacts to the electrodes and the electron-photon interaction (Supporting Information). Depending on the photon energies (infrared to ultraviolet) and electron-photon coupling strength, the switching times required for a smooth relaxation vary in the (sub-)picosecond timescale. All time-resolved calculations are performed with the \textsc{cheers} code~\cite{pavlyukh_cheers_2023}.

\paragraph{Nonzero net currents with zero net driving.}
The molecular junction is driven out of equilibrium by a biharmonic bias profile
\begin{equation}\label{eq:bias}
V_\alpha(t) = V_\alpha^0 + A_\alpha^{(1)}\cos(\Omega_\alpha t + \phi_\alpha) + A_\alpha^{(2)}\cos(\Omega_\alpha t),
\end{equation}
where a constant dc value $V_\alpha^0$ is modulated by first and second harmonics with amplitudes $A_\alpha^{(1)}$ and $A_\alpha^{(2)}$. Both harmonics share the same driving frequency $\Omega_\alpha$, and the phase $\phi_\alpha$ creates a temporal shift between them and the separate electrodes $\alpha\in\{L,R\}$. The bias is relative to the chemical potential $\mu=-1.93$~eV, set in the middle of the HOMO-LUMO gap $\Delta=5.88$~eV. All calculations are done at room temperature $293$~K corresponding to $\beta=40$~eV$^{-1}$. To simplify the large parameter space of the driving protocol in Eq.~\eqref{eq:bias}, we take $V_L^0=V_R^0 \equiv V_0$, $A_L^{(1)}=A_L^{(2)}=A_R^{(1)}=A_R^{(2)}=V_0/2$, and $\Omega_L=\Omega_R\equiv \Omega$. This setup allows for studying various zero-net drivings by adjusting $V_0$ and $\phi\equiv\phi_L-\phi_R$.

We define a \emph{rectified} net electric current by the condition
\begin{equation}\label{eq:elcurr}
\overline{I_{\mathrm{el}}} \equiv \frac{1}{W}\int_0^{t_{\mathrm{f}}} d t w(t) \left[ I_R(t) - I_L(t) \right] \neq 0,
\end{equation}
with $t_{\mathrm{f}}$ being the final simulation time, and $I_\alpha(t)$ are the individual time-dependent currents at the $\alpha$th electrode evaluated from the Jauho-Meir-Wingreen formula~\cite{meir_landauer_1992, jauho_time-dependent_1994, tuovinen_time-linear_2023}
\begin{align}\label{eq:Ia}
I_{\a}(t)=2s_{\a}(t)\Re\Tr\Big[ & \G_{\a}\big(s_{\a}(t)\frac{2\rho(t)-1}{4}\nonumber \\
&-\sum_{\ell}\frac{\h_{\ell}}{\b}\callG^{\rm em}_{\ell\a}(t)\big)\Big].
\end{align}
We average the current oscillations by a Gaussian window $w(t)=\exp[-\frac{(t-t_0)^2}{2W^2}]$ centered around $t_0$ and normalize the current with the spread $W$.

Similarly, the \emph{electroluminescence rectification} is defined by the condition
\begin{equation}\label{eq:ptflux}
\overline{J_{\mathrm{pt}}} \equiv \frac{1}{W}\int_0^{t_{\mathrm{f}}} d t w(t) J(t) \neq 0,
\end{equation}
where the photon energy flux is given by the rate of change of photon number $J(t)\equiv \frac{d}{dt}\sum_{\bgm} \omega_{\boldsymbol{\mu}} n_{\bgm}(t)$. In Ehrenfest approximation the number of photons in each mode is $n_{\bgm}(t)=\sum_\xi\langle\hat{a}_\mu^\dagger(t)\hat{a}_\mu(t)\rangle=\tfrac12\sum_\xi \phi_{\bgm,\xi}^2(t)$. Here, we also average the current oscillations by the Gaussian window, $w$, and then Eq.~\eqref{eq:ptflux} is conveniently performed via integration by parts.

\begin{figure}[t]
 \includegraphics[width=0.475\textwidth]{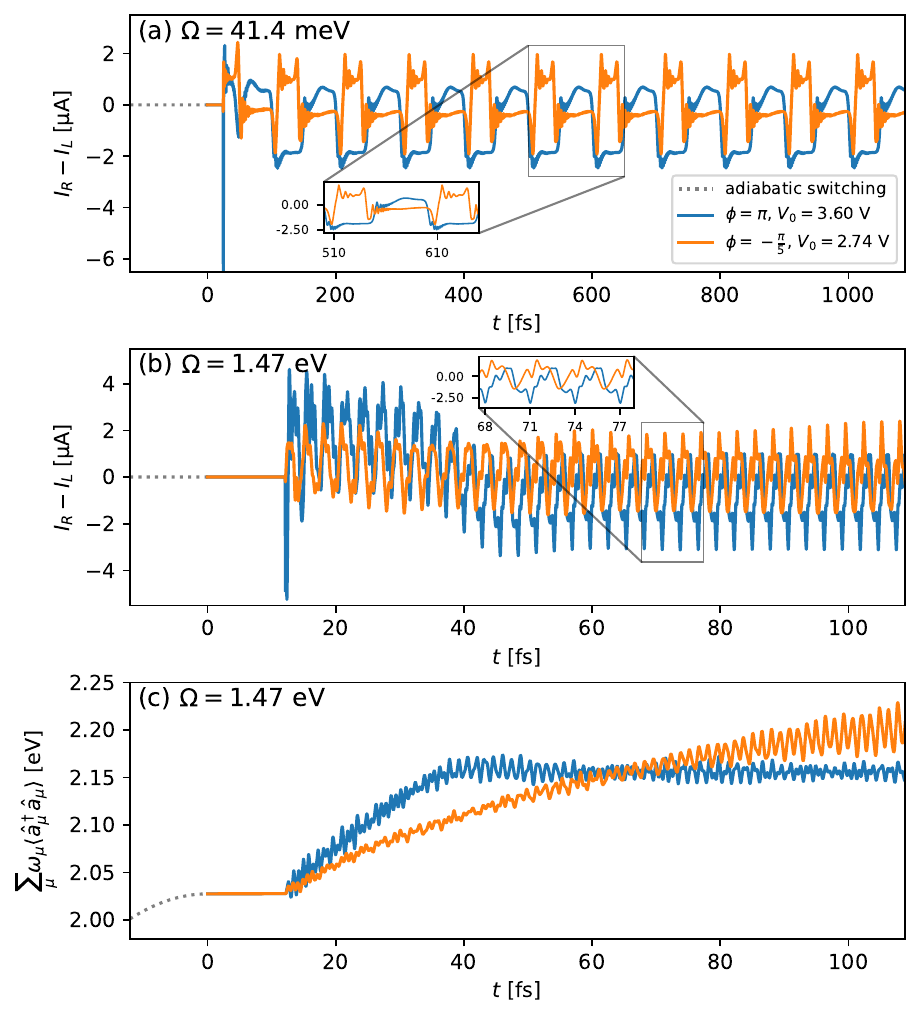}
  \caption{\small Time evolution of (a-b) net electric current and (c) total photon energy for two representative points in the ac driving protocol: $V_\a(t) = V_0 + V_0 \cos(\Omega t + \phi_\a)/2 + V_0 \cos(\Omega t)/2$ where $V_0$ is the constant dc value and $\phi\equiv \phi_L-\phi_R$ is the total phase difference. Adiabatic switching (extending over the figure frame to negative times) is shown with dotted lines. In panel (a) a bias voltage with driving frequency $\Omega=4.14$~meV $=10$~THz ($0.00152$~a.u.) is switched on at $t=24$~fs ($1000$~a.u.). In panels (b-c) a bias voltage is switched on at $t=12$~fs ($500$~a.u.) with frequency $\Omega=1.47$~eV ($0.054$~a.u.). The legend applies to all panels.}
  \label{fig:evolution}
\end{figure}

\begin{figure*}[t]
 \includegraphics[width=\textwidth]{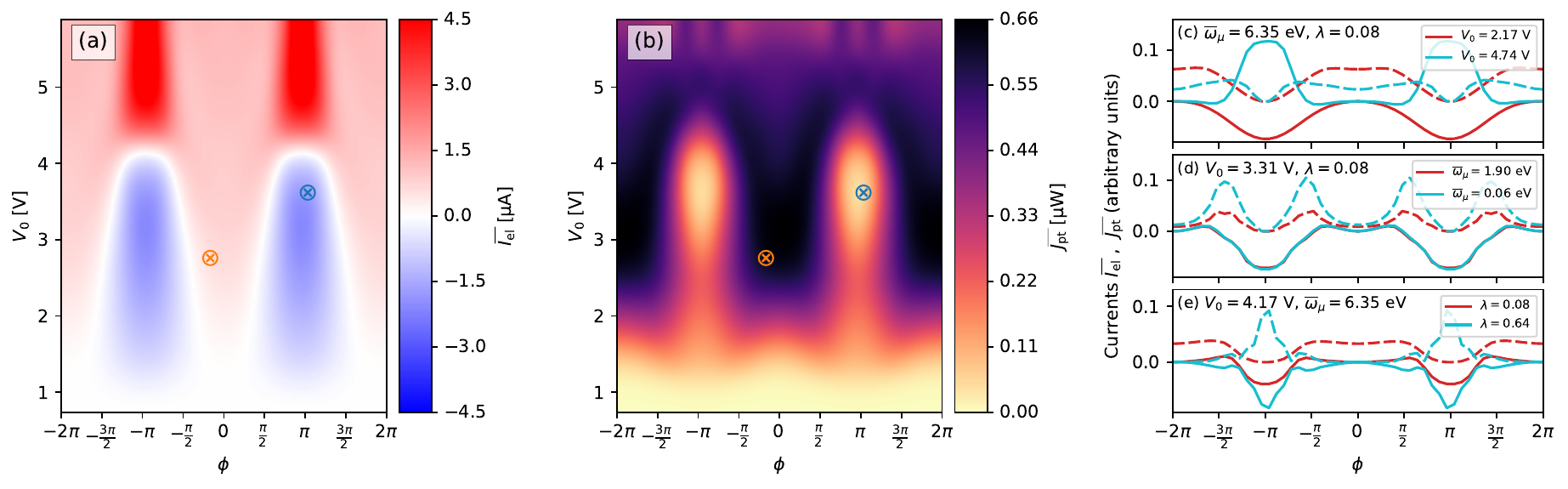}
  \caption{\small Rectified electric current (a) and photon energy flux (b) in terms of the ac driving phase shift $\phi$ (horizontal axes) and central value $V_0$ (vertical axes). The crosses display the two representative time evolutions in Fig.~\ref{fig:evolution} with respective coloring. Panel (c) shows a line cut of panels (a) and (b) for two voltages $V_0$, and panels (d) and (e) show similar line cuts for varying photon energies ($\omega_{\boldsymbol{\mu}}$) and coupling strengths (effective interaction $\lambda$), respectively, cf.~Supporting Information. In panels (c-e), the electric current $\overline{I_{\mathrm{el}}}$ is shown with solid lines and the photon energy flux $\overline{J_{\mathrm{pt}}}$ with dashed lines; the vertical axes are shifted in order to display both electric and photonic currents in the same frame (arbitrary units).}
  \label{fig:rectification-uv-weak-fast}
\end{figure*}

In addition to the above characterization of the molecular junction, we focus on ultraviolet photons, $\omega_{\boldsymbol \mu}=(5.44, 5.44, 8.16)$~eV (mean photon energy $\overline{\omega}_{\boldsymbol{\mu}}=6.35$~eV), and electron-photon coupling of $g=0.218$~eV, classified as the weak-coupling regime by the effective interaction $\lambda\equiv g^2/(\overline{\omega}_{\boldsymbol \mu}\|\Gamma\|) = 0.08$, where $\|\Gamma\|=0.097$~eV denotes the euclidean norm of the total line-width matrix. 
This further indicates a weak-contacting regime (electrode-molecule), where WBLA is a good approximation~\cite{zhu_time-dependent_2005, verzijl_applicability_2013, covito_transient_2018}.
Intermediate ($\lambda=0.43$) to strong ($\lambda=0.64$) coupling with visible-light and infrared photons are detailed in the Supporting Information. Fig.~\ref{fig:evolution} shows the time evolution of net electric current and total photon energy for two driving frequencies: one achievable with state-of-the-art electronics~\cite{urteaga_inp_2017}, $\Omega=41.4$~meV $=10$~THz ($0.00152$~a.u.), and one for systematic parameter investigation, $\Omega=\Delta/4=1.47$~eV ($0.054$~a.u.). The latter allows shorter propagation times to extract rectified currents. We use an adiabatic switching protocol and confirm the observables are saturated before switching on the driving. Phase difference, $\phi=\{-\pi/5,\pi\}$, and voltage center values, $V_0=\{2.74, 3.60\}$~V, illustrate cases of suppressed and enhanced rectification, showing dynamical response and time-varying current due to persistent bias oscillations (periods of $2\pi/\Omega$). A distinct ``ringing'' oscillation~\cite{jauho_time-dependent_1994, ridley_current_2015}, originating from an intra-molecular transition, is also visible causing current signal resonances [insets of Fig.~\ref{fig:evolution}(a,b)]. At periodic stationary states ($t\gtrsim 50$~fs), the photonic flux slope decreases when electric current is most rectified (and vice versa). Comparing slow and fast driving scenarios, similar peak currents are achieved in the range of microamperes (typical for molecular electronics~\cite{matsuhita_conductance_2013,chen_from_2021}), with rectified current magnitude being sensitive to driving frequency. In the following, we consider the case of fast driving, which enhances the rectification effect without affecting the underlying physical mechanisms.

A sweep over the ac driving parameters $\phi$ and $V_0$ is performed, and in Fig.~\ref{fig:rectification-uv-weak-fast}(a-b) we show the rectified electric current and the photon energy flux. These are evaluated from Eqs.~\eqref{eq:elcurr} and~\eqref{eq:ptflux} with the Gaussian window with centering $t_0=73$~fs and spread $W=15$~fs [positioned within the periodic stationary regime; see Fig.~\ref{fig:evolution}(b,c)]. The two representative time evolutions of Fig.~\ref{fig:evolution} are marked by crosses with respective coloring in Fig.~\ref{fig:rectification-uv-weak-fast}(a-b). Key observations are:
1) If the bias is well below the transport channels relative to the Fermi level (HOMO/LUMO), there is no transport or radiation.
2) When ac driving nearly covers individual levels in the transport window but remains below them, a negative rectified current appears. Maximum electric current, at $\phi=\pm\pi$, coincides with minimum photon flux, and vice versa at $\phi=\{0,\pm 2\pi\}$. This is due to ``plasma oscillations'' within the junction: maximal charge sloshing without transmission maximizes radiation, while maximal rectified current reduces oscillations and radiation.
3) When the bias window exceeds the HOMO-LUMO gap, the current maximizes in the positive direction. Unlike single-level transport explained by electron- and hole-transfer processes~\cite{ridley_time-dependent_2017,ridley_many-body_2022}, intramolecular transitions in the junction influence the photon energy flux over various phase differences $\phi$, eliminating minima in the radiation profile.

Electroluminescence rectification also occurs for intermediate to strong couplings and for visible-light and infrared photons. Fig.~\ref{fig:rectification-uv-weak-fast}(c-e) touches on these features, and more detailed data, such as strength dependence of the photon flux on the photon energies and the effective interaction, is available in the Supporting Information. Fig.~\ref{fig:rectification-uv-weak-fast}(c) presents line cuts from Fig.~\ref{fig:rectification-uv-weak-fast}(a-b) for two voltages, fixed ultraviolet photon energy, and weak coupling, showing consistent behavior with Fig.~\ref{fig:rectification-uv-weak-fast}(a-b). Fig.~\ref{fig:rectification-uv-weak-fast}(d) explores different photon energies, $\omega_{\boldsymbol{\mu}}=(1.63, 1.63, 2.45)$~eV (visible light, mean photon energy $\overline{\omega}_{\boldsymbol{\mu}}=1.90$~eV) and $\omega_{\boldsymbol{\mu}}=(54.4, 54.4, 81.6)$~meV (infrared, mean photon energy $\overline{\omega}_{\boldsymbol{\mu}}=0.06$~eV), at fixed voltages and coupling strength. While the electric current response remains similar, the photon flux varies, particularly with infrared photons, where additional maxima around $\phi=\pm \pi/2$ and $\phi=\pm 3\pi/2$ appear due to `slower' cavity photons forming metastable configurations. Fig.~\ref{fig:rectification-uv-weak-fast}(e) demonstrates that strong electron-photon coupling can completely reverse the electroluminescence rectification, suggesting potential breakdowns of HF+Ehrenfest dynamics or competing mechanisms at different time scales, which shall be addressed next.

\begin{figure*}[t]
\centering
\includegraphics[width=0.97\linewidth]{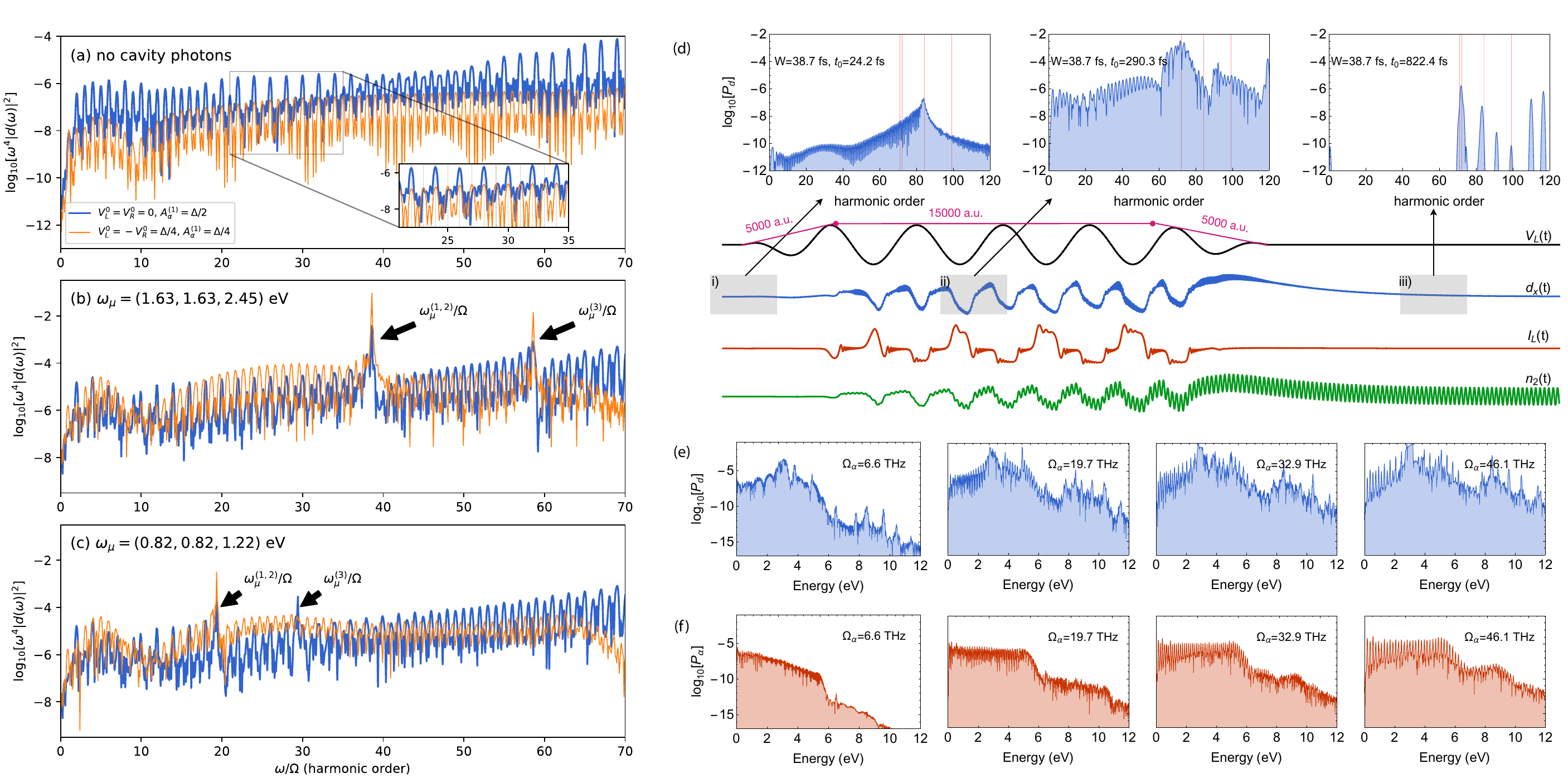}
  \caption{ \small HHG power spectrum of the molecular junction evaluated by the Fourier transform of the dipole moment. (a) Without cavity photons, (b) weak coupling to visible-light photons, (c) weak coupling to infrared photons. Solid blue lines denote the first bias protocol while thin orange lines denote the second bias protocol. For the same set of parameters as in (c), the dependence of the dipole power spectrum on the position of the window is demonstrated in panel (d), whereas in panels (e) and (f) the dependence of the dipole moment and electric current power spectra on the driving frequency $\Omega$ is depicted. The three small panels in (d) represent the instantaneous Fourier transform of the dipole moment within specific time windows indicated in the $d_x(t)$ plot [using a Gaussian window, cf. Eqs.~\eqref{eq:elcurr} and~\eqref{eq:ptflux}]. Red vertical lines mark the molecular excited states, computed for the isolated molecule by solving the Casida equation~\cite{casida_time-dependent_2009} within the time-dependent Hartree-Fock framework (without photons).}
  \label{fig:hhg}
\end{figure*}

\paragraph{High-harmonic generation.}
Our simulations demonstrate that the molecular junction exhibits pronounced high-harmonic generation due to their non-linear characteristics (cf.~Eqs.~\eqref{eq:rho:eom}-\eqref{eq:gla:eom}). Importantly, by modifying the external bias we uncover phenomena that cannot be obsered in conventional optical setups, e.g., enhance or suppress the even harmonics response. To simplify the discussion, we concentrate on a monoharmonic drive: In Eq.~\eqref{eq:bias}, set $A_\alpha^{(2)}=0$. It can be expected that bias profiles with odd inversion symmetry, $V_\alpha(t+\pi/\Omega)=-V_\alpha(t)$, tend to suppress even harmonics, whereas breaking this symmetry enhances them~\cite{ridley_time-dependent_2017,tuovinen_time-resolved_2019}. This is demonstrated by choosing $V_L^0=V_R^0=0$, $A_\alpha^{(1)}=\Delta/2$ and $\phi=\pi$ for the former case, and $V_L^0=V_R^0=\Delta/4$, $A_\alpha^{(1)}=\Delta/4$ and $\phi=\pi$ for the latter, asymmetric case. We set the driving frequency by the slower protocol of Fig.~\ref{fig:evolution}(a), $\Omega=41.4$~meV $=10$~THz ($0.00152$~a.u.).

In Fig.~\ref{fig:hhg}, we present the high-harmonic generation (HHG) power spectrum
\begin{align}
P_d(\w)&=\w^4\sum_{p=x,y,z} |\smallint d_p(t) e^{i\w t} dt|^2
\end{align}
as the Fourier transform of the electric dipole moment
\begin{align}
d_p(t)&=\sum_{i,j}d_{p,ij} \rho_{ji}(t),\label{eq:dipole}
\end{align}
using the matrix element $d_{p,ij}$ of the $p$th projection of the molecule's dipole operator. We employ a trapezoidal envelope to prevent excitation of high-energy modes by sudden voltage increases [Fig.~\ref{fig:hhg}(d)]. The spectrum shows suppression of even harmonic multiples in odd inversion symmetric driving within certain frequency ranges. The inset of Fig.~\ref{fig:hhg}(a) displays distinct peaks at frequencies $\omega=(2n+1)\Omega$ with $n$ integer. In contrast, with asymmetric driving, peaks appear at even multiples, $\omega=2n\Omega$, of the driving frequency. This selection between even/odd harmonics is very clearly present in the electric current signal (Supporting Information), which is also partially visible when interacting with cavity photons, as shown in panels (b) and (c).

Intrinsic molecular excitations are imprinted in the HHG spectra. Notably, the two distinct peaks in Fig.~\ref{fig:hhg}(b,c) linked to photon cavity modes and their strong interaction with the electric dipole, clearly shift with varying photon energy. Higher-energy peaks (harmonic order $\gtrsim 80$) correlate with intramolecular transitions, as detailed in the time- and energy-resolved HHG spectra in Fig.~\ref{fig:hhg}(d). Initially, an electronic state at $3.49$~eV with high oscillator strength dominates, but as the pulse progresses, energy is redistributed and other intramolecular states are excited. Additionally, the spectral peaks are slightly red-shifted and broadened compared to the isolated molecule. Further simulations rescaling the $\Gamma_\alpha$-matrices (not shown) confirmed these effects are related to the embedding correlator of the electrodes.

In Figs.~\ref{fig:hhg}(e) and~\ref{fig:hhg}(f), we examine how driving frequency affects the high-harmonic dipolar and electric current spectra. The overall HHG signal shows weak dependence on driving frequency. The current response displays a two-plateau shape typical in optical HHG experiments, but the cut-off energy does not strongly depend on excitation frequency. The explanation is similar to HHG in bulk materials: the three-step model~\cite{schafer_above_1993, corkum_plasma_1993}, valid for gas systems assuming free electron evolution, loses validity~\cite{tancogne-dejean_atomic-like_2018}. In solids, electrons do not behave as free particles due to non-parabolic band structures~\cite{tancogne-dejean_impact_2017}. In our model of the molecular junction, electronic reservoirs are treated in the wide band limit, and electron motion in the electrodes is neglected, which is physically justified and distinct from the free-electron electrodes. Within GKBA, anything but WBLA would be inconsistent and could lead to a finite recurrence time-scale for the leads and spurious physical effects~\cite{tuovinen_time-linear_2023}. Thus, HHG from molecular junctions is more akin to that from solids than from isolated molecular systems.

We emphasize that bias-driven molecular junctions are more complex than typical atomic, molecular, or bulk systems: HHG is simultaneously observed in dipolar radiation, electric current, and photonic flux, with very high harmonic orders in all cases. This shows strong upconversion in electrically-driven photoemission, where generated photon energies far exceed the applied bias, indicating operation well beyond the linear-response regime.

\paragraph{Conclusion.}
We studied electrically-driven molecular junctions, focusing on electroluminescence and the rectification or pumping of electrical currents and photon fluxes. We showed that coupling electrons within the molecule to cavity photons creates a complex nonequilibrium scenario, necessitating equal treatment of electron-electron and electron-boson interactions, external driving, and coupling to metallic electrodes. Using a time-linear, wide-band limit NEGF approach for a realistic molecular junction required developing an atomistic formalism, partitioning the system into a central fragment and electrodes, and determining the effective Hamiltonian and tunneling matrices.

We focused on Benzenedithiol molecular junctions driven by periodic fields in both transient and stationary regimes. Using a fully atomistic all-electron approach, we performed geometry optimization at the hybrid DFT level. This approach, while neglecting some correlations, aligns with studies involving larger electrodes~\cite{silva_structural_2020}. Higher-order electronic and photonic correlations, though feasible within our scheme~\cite{pavlyukh_interacting_2022, tuovinen_time-linear_2023}, are beyond the excitation regimes considered.
Employing a biharmonic bias profile, we demonstrated how phase differences between oscillating voltages affect long-time rectified electric and photonic currents. We also calculated the electric dipole moment response to monoharmonic driving. By adjusting the inversion symmetry of the ac bias profile, we selectively generated high-order harmonics of odd or even multiples of the driving frequency. Our findings imply that discrete on/off states of electric and photonic currents could be produced from analog waves, which is significant for designing molecular-junction-based frequency modulators, switches, and detectors~\cite{mittleman_perspective_2017, wang_modulation_2017, hnid_molecular_2024}. Our current values align with steady-state results from Ref.~\citenum{silva_structural_2020} and experimental values in Ref.~\citenum{kim_benzenedithiol_2011}. Comparing the power carried by the photon flux [Fig.~\ref{fig:rectification-uv-weak-fast}(b)] with the total power, we found quantum efficiency around ten percent (Supporting Information), similar to quantum-dot devices~\cite{yang_high-efficiency_2015,won_highly_2019,meng_ultrahigh-resolution_2022}.

The presented methodology holds promise for future studies on ab initio time-resolved quantum transport in molecular junctions. We foresee a potentially illuminating, direct extension of the present work to the time-dependent current fluctuations and the associated noise spectra~\cite{kaasbjerg_theory_2015, cabra_local-noise_2018, ridley_electron_2019, cohen_greens_2020, tikhonov_spatial_2020, ridley_many-body_2022} of quantum correlated molecular junctions out of equilibrium.

Supporting Information. Details of the numerical simulations, electronic and photonic rectification profiles, time-dependent electric currents and photonic fluxes, and dipolar responses in the HHG setting.

\begin{acknowledgement}
Y.P. acknowledges funding from the project No. 2021/43/P/ST3/ 03293 co-funded by the National Science Centre and the European Union Framework Programme for Research and Innovation Horizon 2020 under the Marie Sklodowska-Curie grant agreement No. 945339. We also acknowledge CSC--IT Center for Science, Finland, for computational resources.
\end{acknowledgement}

\providecommand{\latin}[1]{#1}
\makeatletter
\providecommand{\doi}
  {\begingroup\let\do\@makeother\dospecials
  \catcode`\{=1 \catcode`\}=2 \doi@aux}
\providecommand{\doi@aux}[1]{\endgroup\texttt{#1}}
\makeatother
\providecommand*\mcitethebibliography{\thebibliography}
\csname @ifundefined\endcsname{endmcitethebibliography}
  {\let\endmcitethebibliography\endthebibliography}{}

\begin{figure}[p]
 \includegraphics[width=3.25in]{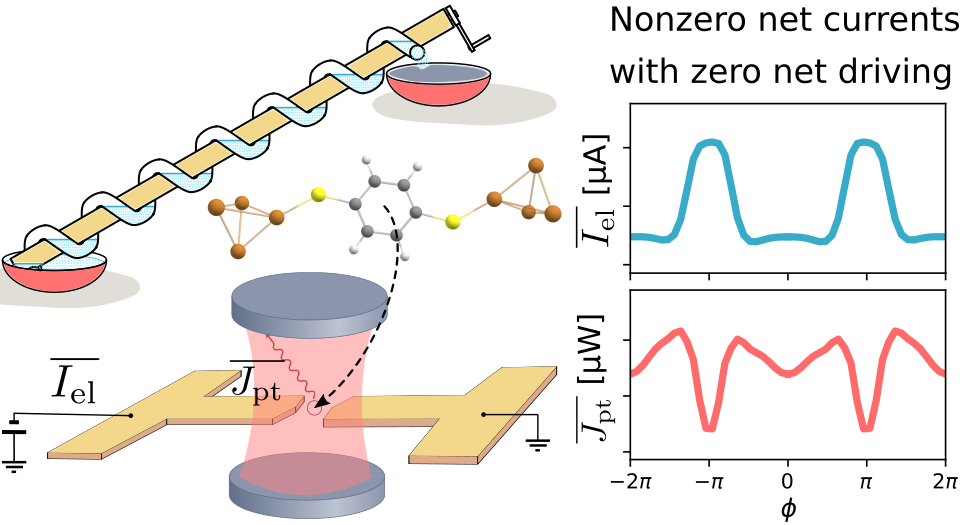}
  \caption{For Table of Contents Only}
  \label{fig:toc}
\end{figure}

\newpage
\onecolumn

\begin{center}
\textbf{\large Supporting Information: Electroluminescence rectification and high harmonic generation in molecular junctions}
\end{center}
%%%%%%%%%% Merge with supplemental materials %%%%%%%%%%
%%%%%%%%%% Prefix a "S" to all equations, figures, tables and reset the counter %%%%%%%%%%
\setcounter{equation}{0}
\setcounter{figure}{0}
\setcounter{table}{0}
\setcounter{page}{1}
\makeatletter
\renewcommand{\theequation}{S\arabic{equation}}
\renewcommand{\thefigure}{S\arabic{figure}}
\renewcommand{\thepage}{S\arabic{page}}
\renewcommand{\thetable}{S\arabic{table}}
\renewcommand{\bibnumfmt}[1]{[S#1]}
\renewcommand{\citenumfont}[1]{S#1}

\paragraph{Preparing a coupled initial state and restarting}

The initial state for our time-dependent simulations is the uncoupled system consisting of the BDT molecular junction.
In order to describe the time-dependent response to, e.g., bias-voltage driving, a preparatory simulation is first needed to reach a coupled (and correlated) equilibrium state. In our simulation protocol, negative times ($t<0$) correspond to this preparatory step, in which we adiabatically switch on the coupling to the leads and many-body interactions (electron-electron and electron-photon) with the same envelope function $s(t)=s_\a(t)=s_\w(t)=\sin^2(\frac{\pi}{2}\frac{t}{t_i})$ for $t\leq t_i$ and $s(t)=1$ otherwise, with $t_i$ being the initial switching time. It needs to be assessed case-by-case that the preparatory step is indeed adiabatic and no spurious transient effects take place due to the switching. In practice this can be done by continuing the simulation for $t\geq 0$ in the absence of external fields (e.g. bias-voltage driving) to verify that typical observables, such as electronic populations and currents remain stable.

In our time-linear framework with the \textsc{cheers} code~\cite{pavlyukh_cheers_2023_supp}, there are no memory integrals (in contrast to the equivalent integro-differential GKBA formalism~\cite{tuovinen_comparing_2020_supp,tuovinen_electronic_2021_supp}). This means we may simply save the adiabatically prepared equilibrium state, and start a new simulation by using this as the initial state. This reduces the computational cost significantly since the same initial state can be utilized for a large number of different out-of-equilibrium simulations.

For the numerical solution of Eqs.~(1-4) in the main text, we employ the fourth order Runge-Kutta (RK4) method, and we use a fixed step length of $\delta t = 0.01$~a.u. $\approx 0.24$~as. We have monitored that the relative error of the RK4 stepping stays below $10^{-6}$.

\paragraph{Electronic and photonic rectification profiles}

Similar to Fig.~4 of main text, here we present auxiliary plots for electric current and photon flux rectification for various cavity-photon energies and electron-photon coupling strengths, see Tab.~\ref{tab:ptparams} and Figs.~\ref{fig:rectification-uv-weak-fast}-\ref{fig:rectification-ir-strong-fast}. The line cuts of Fig.~4(d,e) of main text are obtained from Figs.~\ref{fig:rectification-vis-weak-fast}, \ref{fig:rectification-ir-weak-fast}, and~\ref{fig:rectification-uv-strong-fast}. The coupling strength between electrons and cavity photons is analyzed in terms of the effective interaction $\lambda\equiv g^2/(\overline{\omega}_{\boldsymbol \mu}\|\Gamma\|)$, where $\|\Gamma\|$ denotes the euclidean norm of the total line-width matrix, which in our Cu-BDT-Cu junction is $\|\Gamma\|=0.003574$~a.u. The effective interaction strength thus relates the different energy scales of our setting: How strongly the electrons and photons are coupled together and how fast/slow the dissipation of electrons to the metallic leads is compared to the cavity photons.

\begin{table}[h!]
\centering
\begin{tabular}{c|c|c|c}
& photon energies $\omega_{\boldsymbol{\mu}}$ & el-pt coupling $g$ (a.u.) & eff. interaction $\lambda$ (dim.less)\\
\hline\hline
UV & $(0.2, 0.2, 0.3)$~a.u. & $0.008$ & $0.08$ (weak) \\
   & $\approx(5.44, 5.44, 8.16)$~eV & $0.019$ & $0.43$ (intermediate) \\
   &                   & $0.023$ & $0.64$ (strong) \\
\hline
VIS & $(0.06, 0.06, 0.09)$ & $0.0046$ & $0.08$ (weak) \\
   & $\approx(1.63, 1.63, 2.45)$~eV & $0.0103$ & $0.43$ (intermediate) \\
   &                   & $0.0127$ & $0.64$ (strong) \\
\hline
IR & $(0.002, 0.002, 0.003)$ & $0.0008$ & $0.08$ (weak) \\
   & $\approx(54.4, 54.4, 81.6)$~meV & $0.0019$ & $0.43$ (intermediate) \\
   &                   & $0.0023$ & $0.64$ (strong) \\
\hline
\end{tabular}
\caption{Cavity-photon parameters: ultraviolet (UV), visible-light (VIS), infrared (IR).}
\label{tab:ptparams}
\end{table}

We observe for weak electron-photon interaction, the rectification profiles are qualitatively similar between UV, VIS and IR. For intermediate to strong coupling it is possible to enhance the electroluminescence rectification or even reverse the behaviour of the photon flux, cf.~Fig.~4(e) in the main text.

\begin{figure}[h!]
 \includegraphics[width=0.85\textwidth]{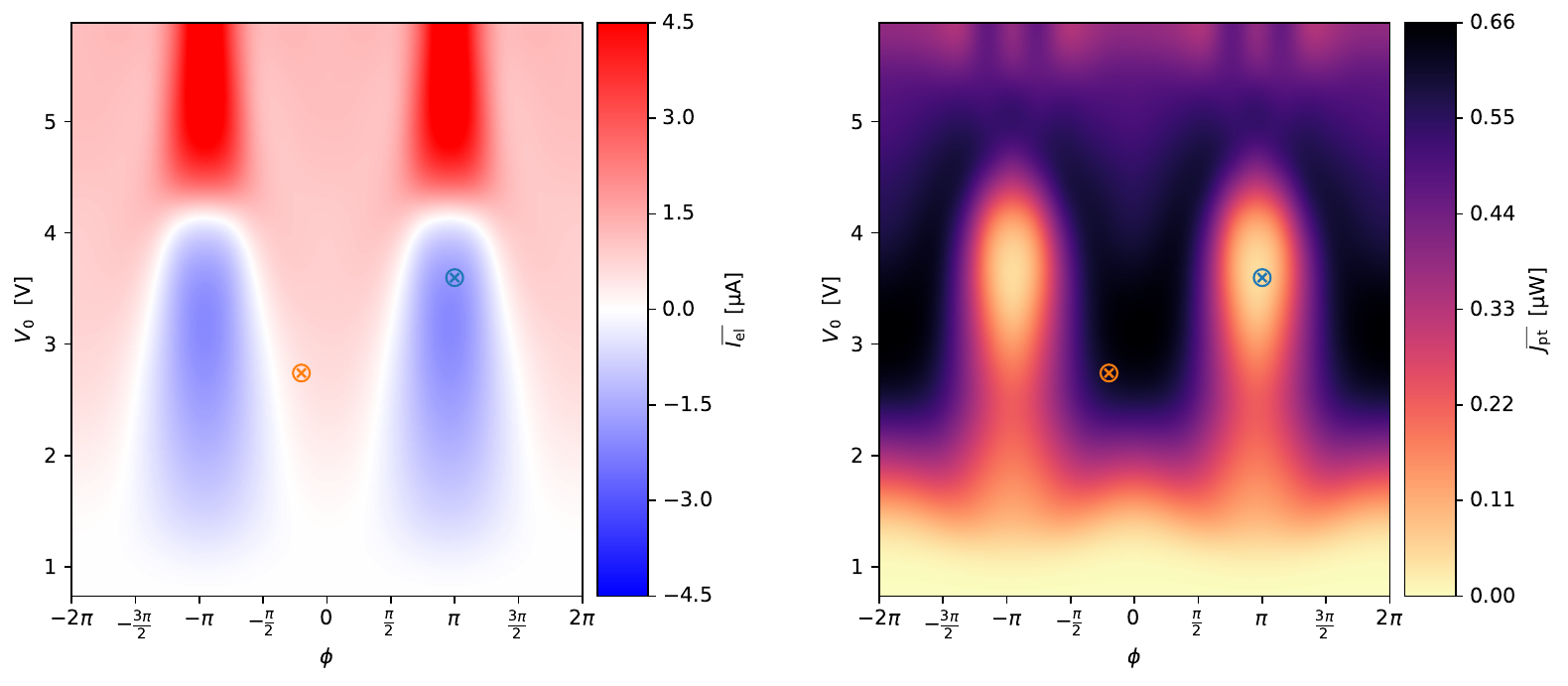}
  \caption{UV photons and weak coupling. This data is also shown in Fig.~4(a,b) in the main text, where the annotated points are discussed.}
  \label{fig:rectification-uv-weak-fast}
\end{figure}

\begin{figure}[h!]
 \includegraphics[width=0.85\textwidth]{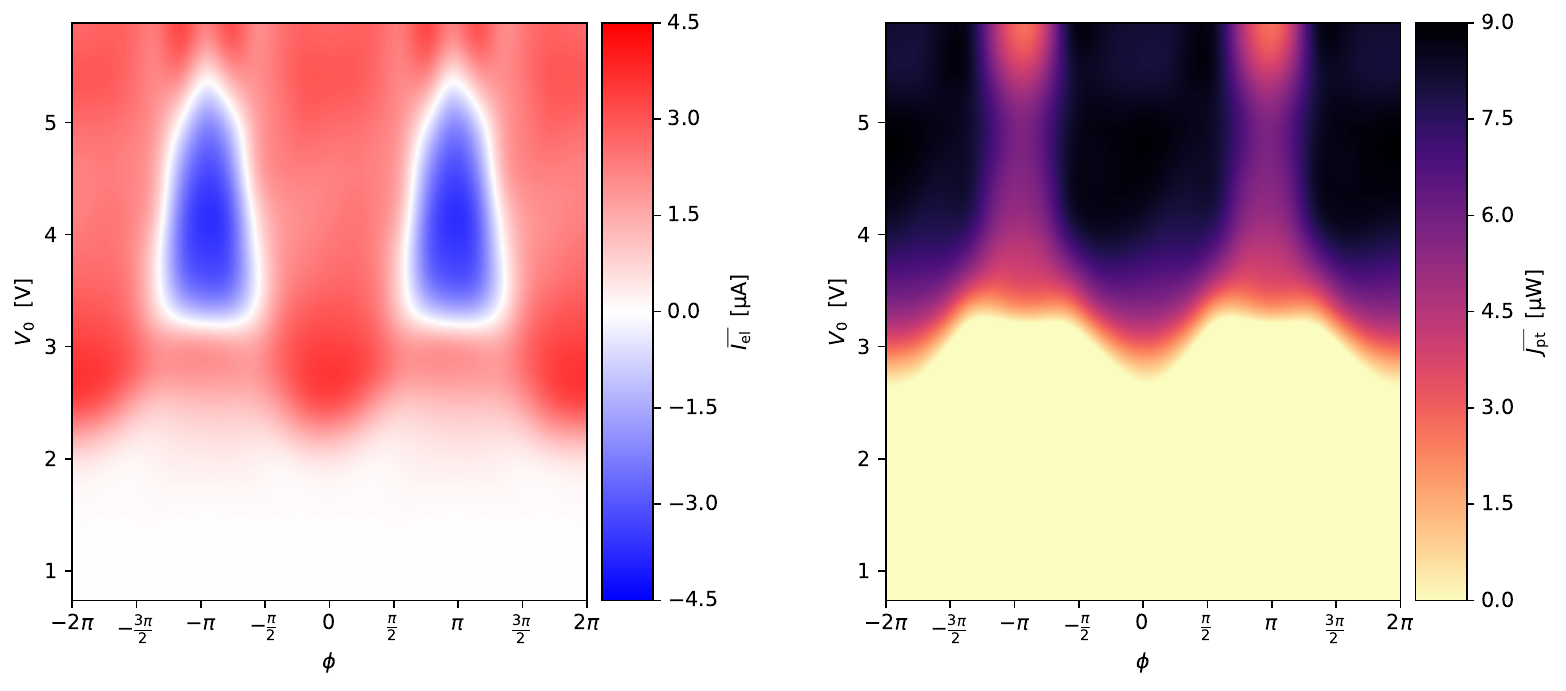}
  \caption{UV photons and intermediate coupling.}
  \label{fig:rectification-uv-intermediate-fast}
\end{figure}

\begin{figure}[h!]
 \includegraphics[width=0.85\textwidth]{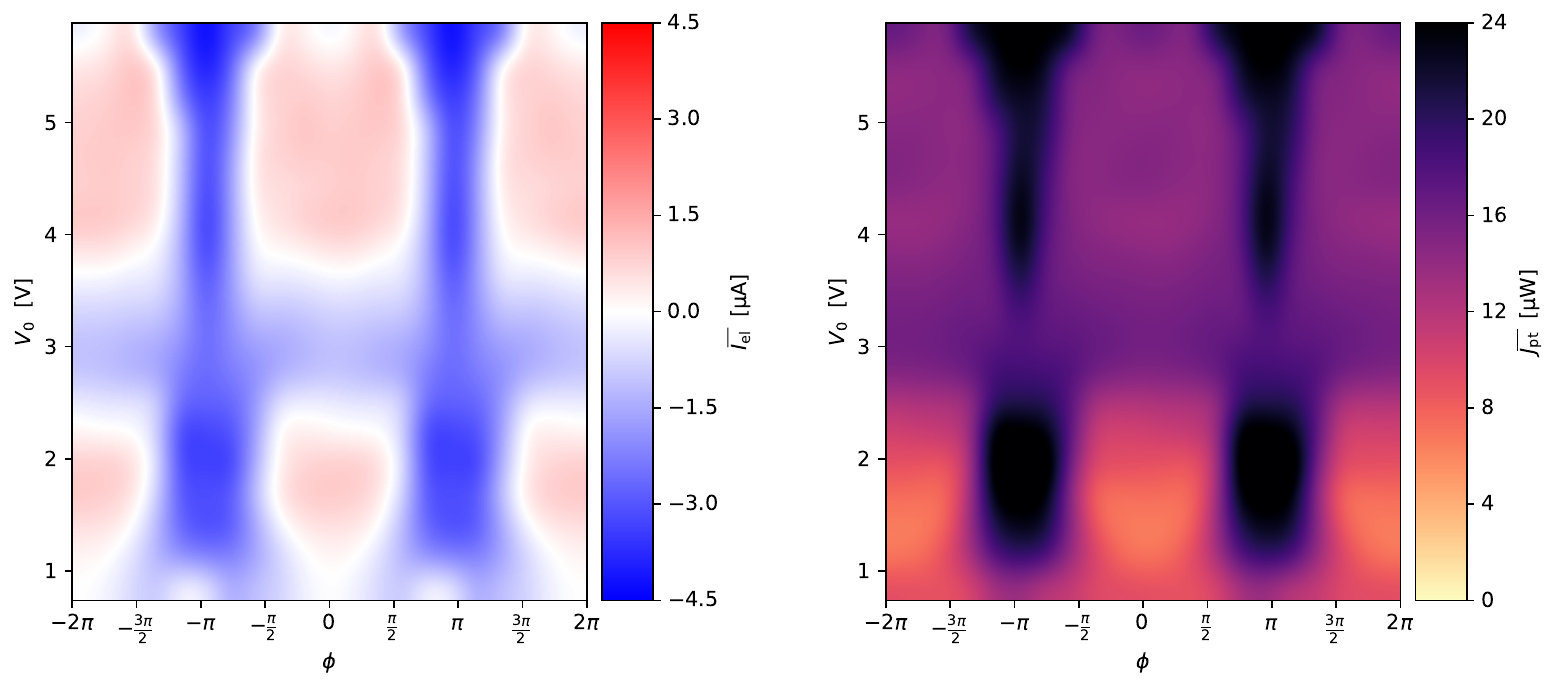}
  \caption{UV photons and strong coupling.}
  \label{fig:rectification-uv-strong-fast}
\end{figure}

\begin{figure}[h!]
 \includegraphics[width=0.85\textwidth]{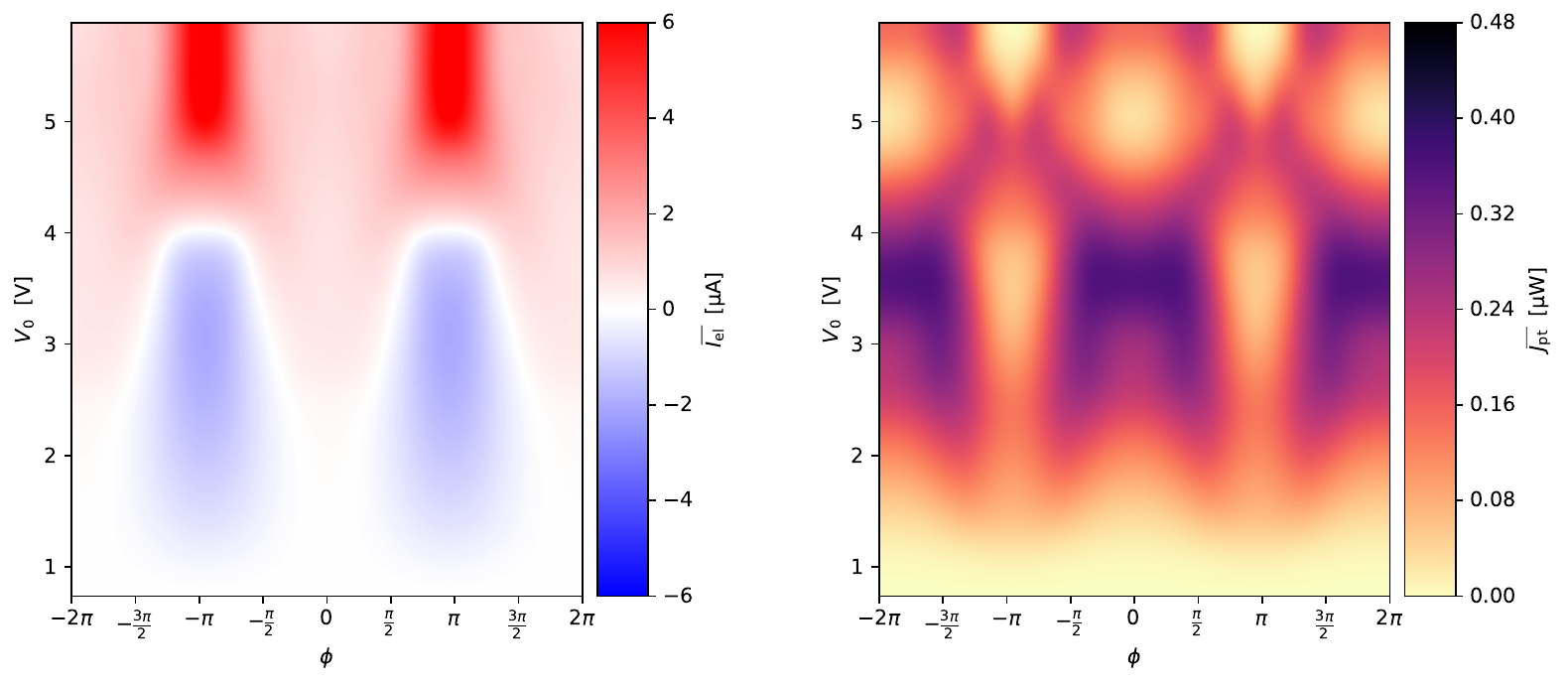}
  \caption{VIS photons and weak coupling.}
  \label{fig:rectification-vis-weak-fast}
\end{figure}

\begin{figure}[h!]
 \includegraphics[width=0.85\textwidth]{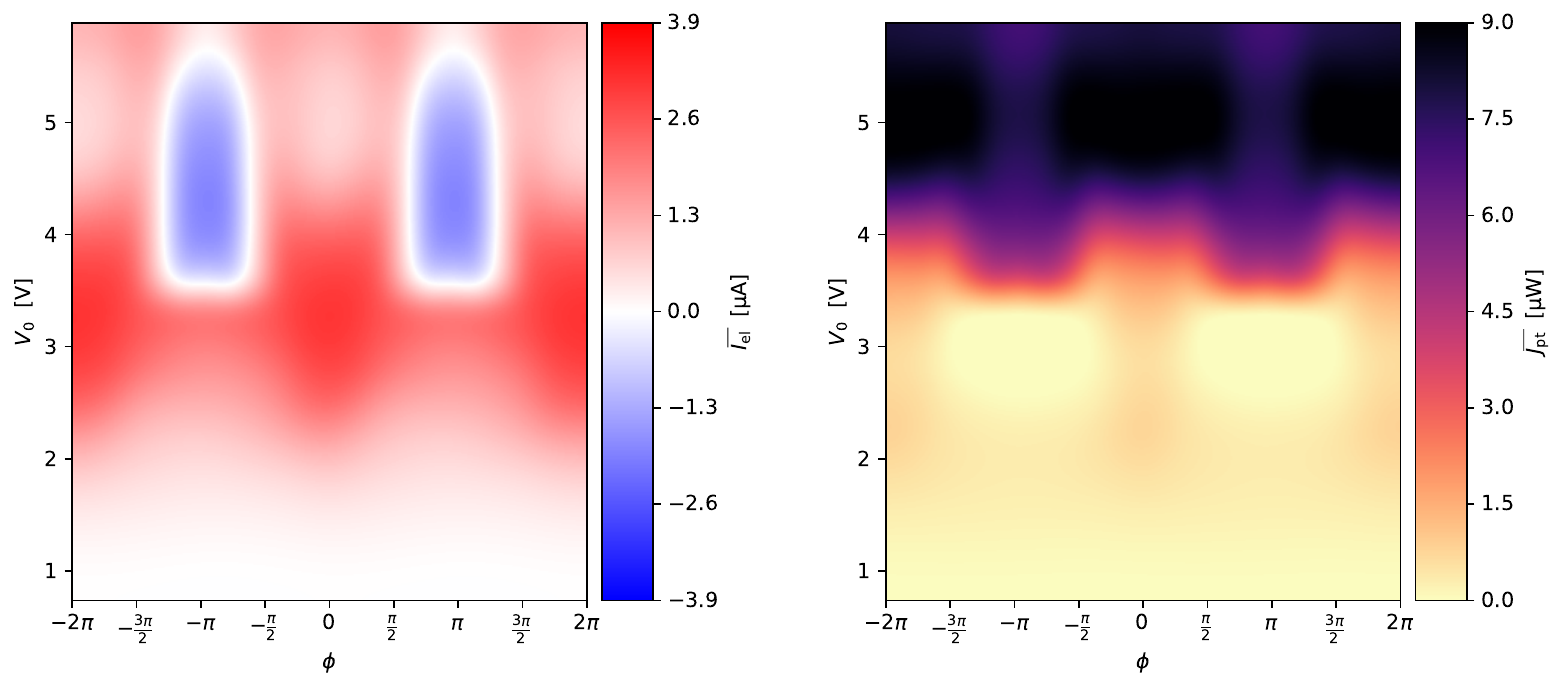}
  \caption{VIS photons and intermediate coupling.}
  \label{fig:rectification-vis-intermediate-fast}
\end{figure}

\begin{figure}[h!]
 \includegraphics[width=0.85\textwidth]{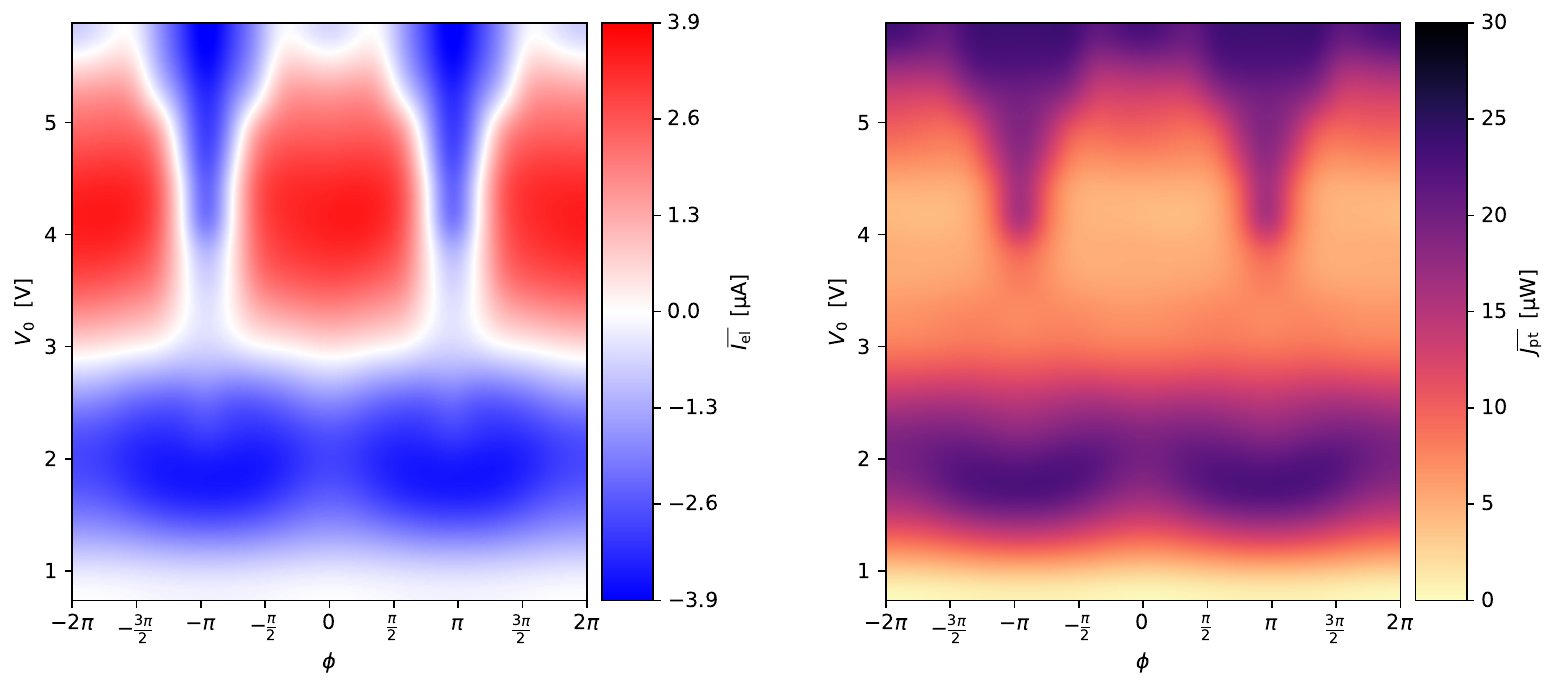}
  \caption{VIS photons and strong coupling.}
  \label{fig:rectification-vis-strong-fast}
\end{figure}

\begin{figure}[h!]
 \includegraphics[width=0.85\textwidth]{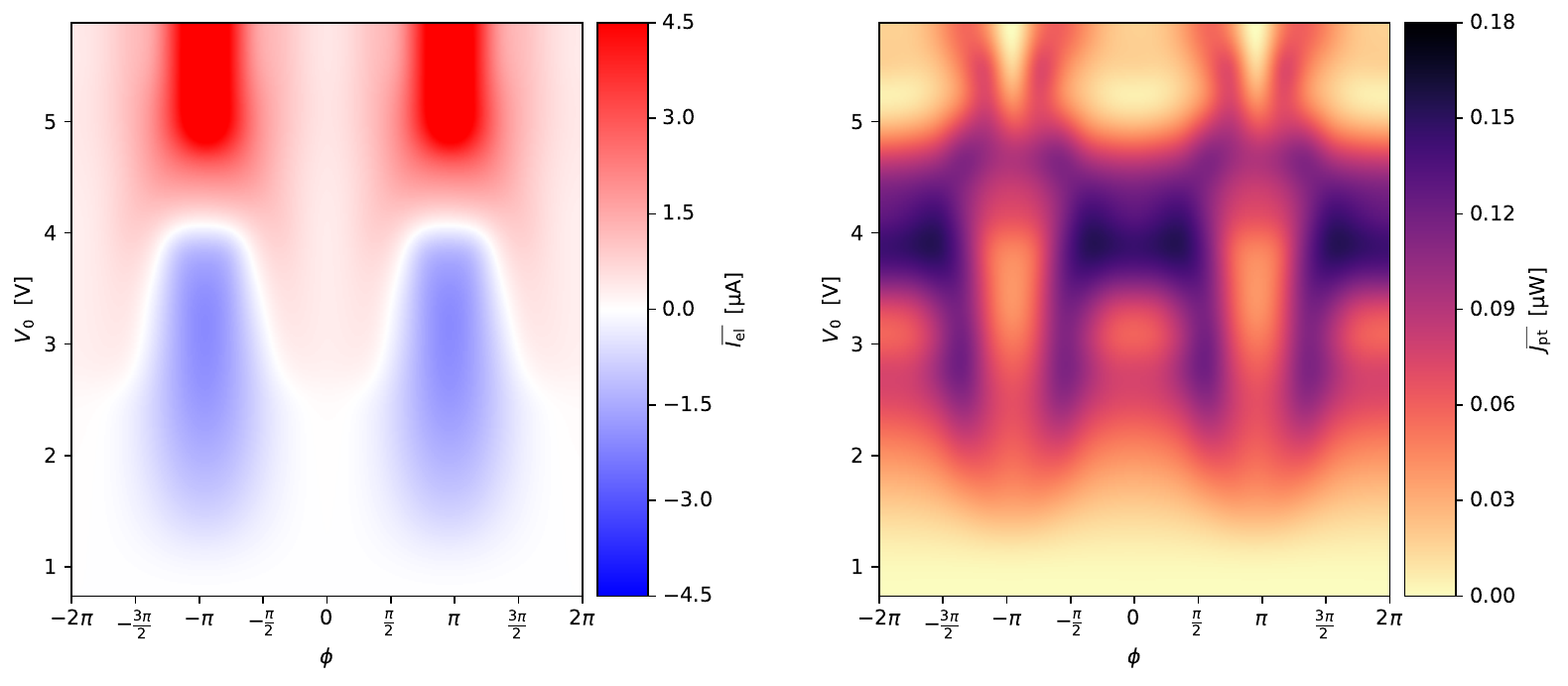}
  \caption{IR photons and weak coupling.}
  \label{fig:rectification-ir-weak-fast}
\end{figure}

\begin{figure}[h!]
 \includegraphics[width=0.85\textwidth]{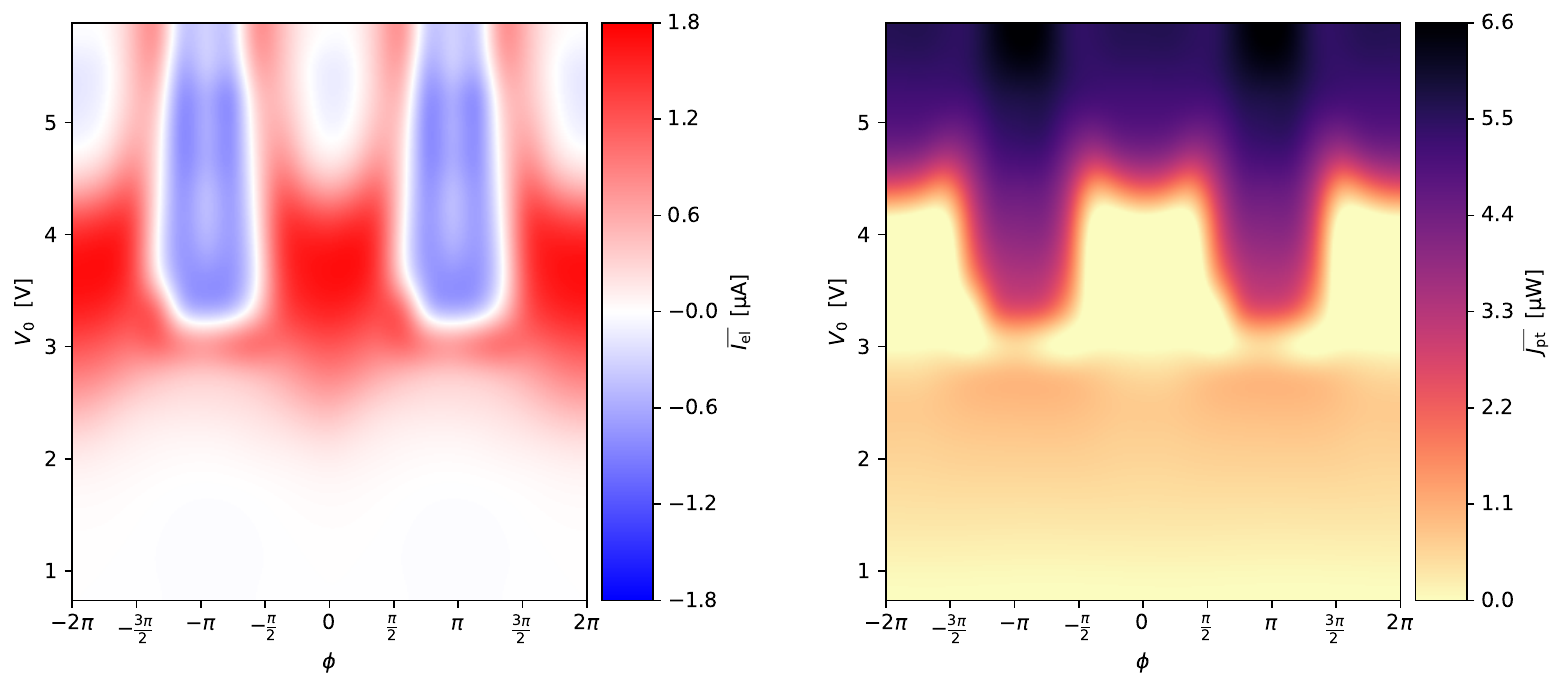}
  \caption{IR photons and intermediate coupling.}
  \label{fig:rectification-ir-intermediate-fast}
\end{figure}

\begin{figure}[h!]
 \includegraphics[width=0.85\textwidth]{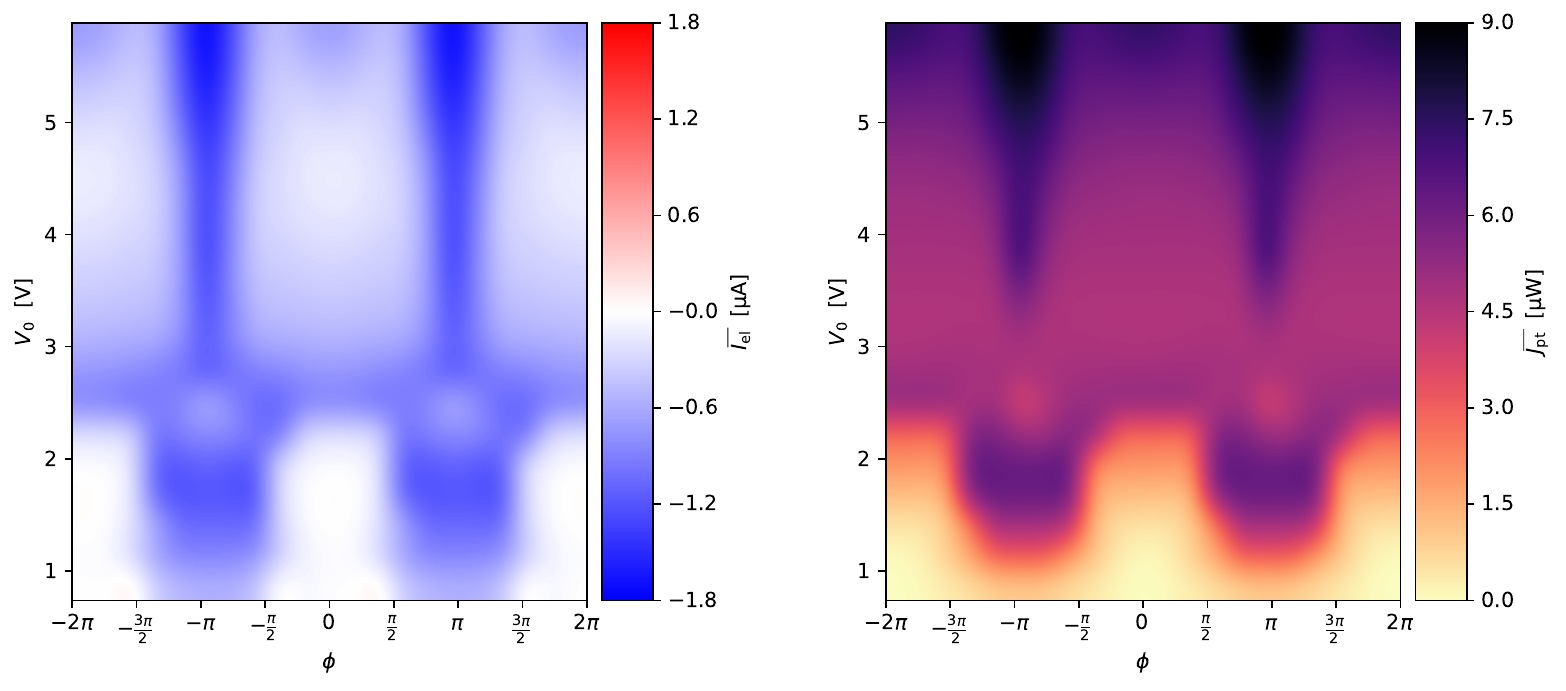}
  \caption{IR photons and strong coupling.}
  \label{fig:rectification-ir-strong-fast}
\end{figure}

Based on the rectification profiles of the photonic energy flux, we may estimate the energy-conversion efficiency of such a molecular-junction based device. For example, in Fig.~\ref{fig:rectification-uv-weak-fast}, the orange cross corresponds to $\overline{I}_{\mathrm{el}}\approx 0{.}55$~µA, which results from an oscillating electric current of maximum value $I_{\mathrm{el}}^{\mathrm{max}}\approx 1{.}88$~µA [cf.~Fig.~3(b) in the main text]. At voltage $V=2.74$~V, this would amount to the total power ${P}_{\mathrm{el}}^{\mathrm{max}} = V{I}_{\mathrm{el}}^{\mathrm{max}} \approx 5{.}15$~µW. The corresponding point in the photon energy flux gives $\overline{J}_{\mathrm{pt}}\approx 0{.}63$~µW, which leads to the efficiency $\overline{J}_{\mathrm{pt}}/{P}_{\mathrm{el}}^{\mathrm{max}} \approx 12\%$. Overall, based on the data reported in Figs.~\ref{fig:rectification-uv-weak-fast}-\ref{fig:rectification-ir-strong-fast} similar comparisons result in the quantum efficiency levels in the neighborhood of ten percent.

\paragraph{Time-dependent response and higher-order harmonics}

In Fig.~5 of the main text, we present high-harmonic generation spectra by using the Fourier transformation of the time-dependent dipole moment. While this corresponds to the experimentally relevant power spectrum, similar characteristics appear also in the electric and photonic currents. We thus define the power spectra of the electric and photonic currents [Eqs.~(10) and~(11) of the main text] as
\begin{align}
P_\a(\w)&=\w^2|\smallint I_\a(t) e^{i\w t}|^2 dt,\\
P_\text{pt}(\w)&=\w^2|\smallint J(t) e^{i\w t} dt|^2.
\end{align}

\begin{figure}[h!]
 \includegraphics[width=0.75\textwidth]{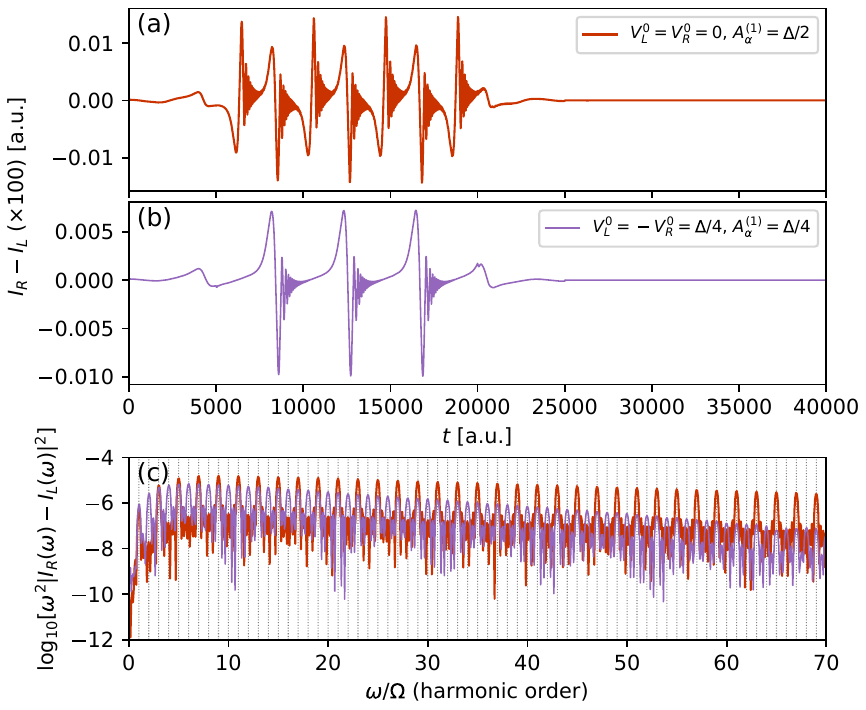}
  \caption{(a-b) Time-dependent electric current response to two monoharmonic bias-voltage driving of frequency $\Omega=0.00152$~a.u., without cavity photons, cf.~Fig.~5(a) in the main text. (c) Fourier transformation of the current signals.}
  \label{fig:hhg-currents-nopt}
\end{figure}

\begin{figure}[h!]
 \includegraphics[width=0.75\textwidth]{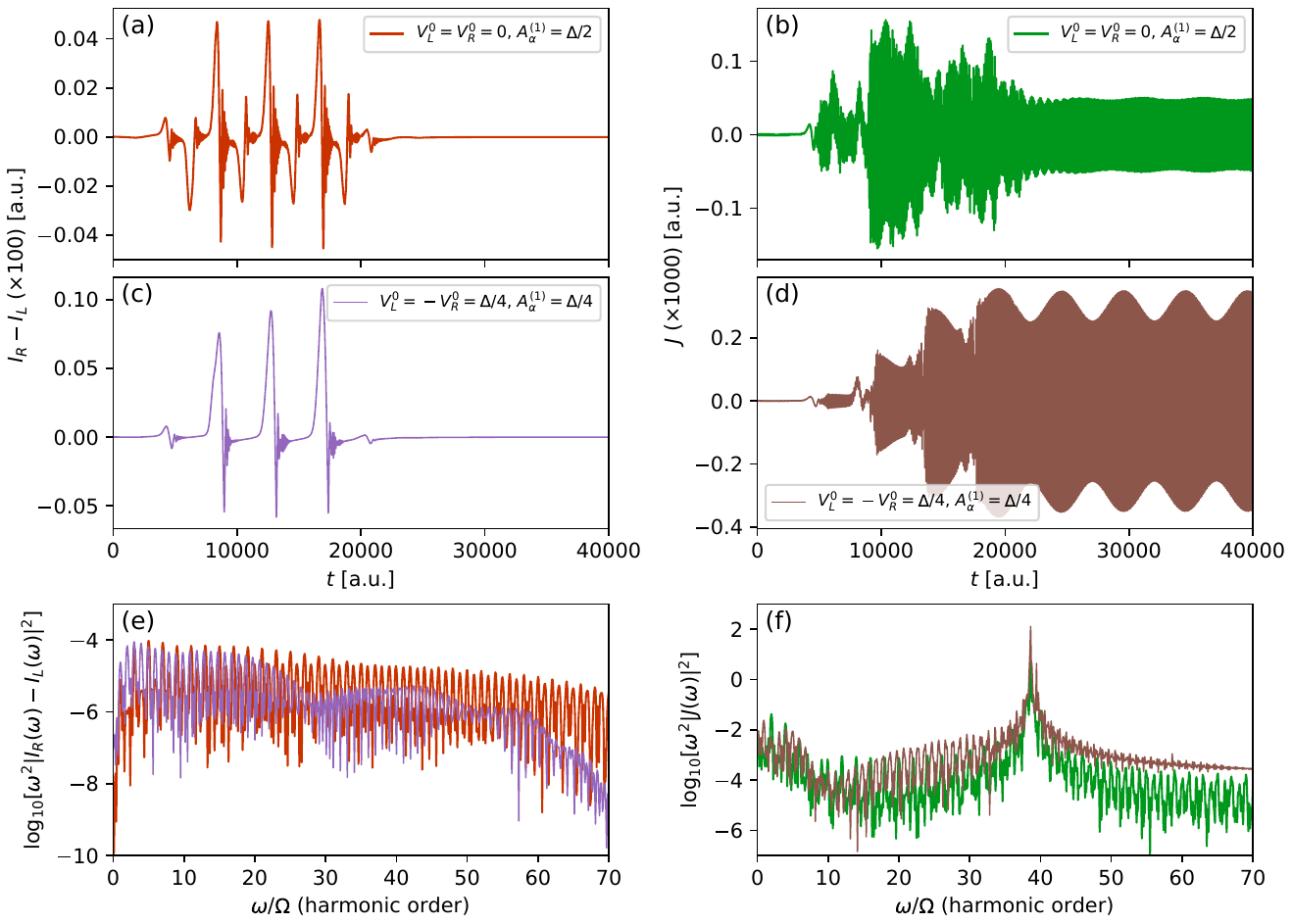}
  \caption{In addition to Fig.~\ref{fig:hhg-currents-nopt}, here we display both electric (a,c) and photonic currents (b,d), and the associated Fourier transformations in panels (e) and (f). The electrons are weakly coupled to cavity photons of energy $\omega_{\boldsymbol{\mu}}=(0.06,0.06,0.09)$~a.u., cf.~Fig.~5(b) in the main text.}
  \label{fig:hhg-currents-vis}
\end{figure}

\begin{figure}[h!]
 \includegraphics[width=0.75\textwidth]{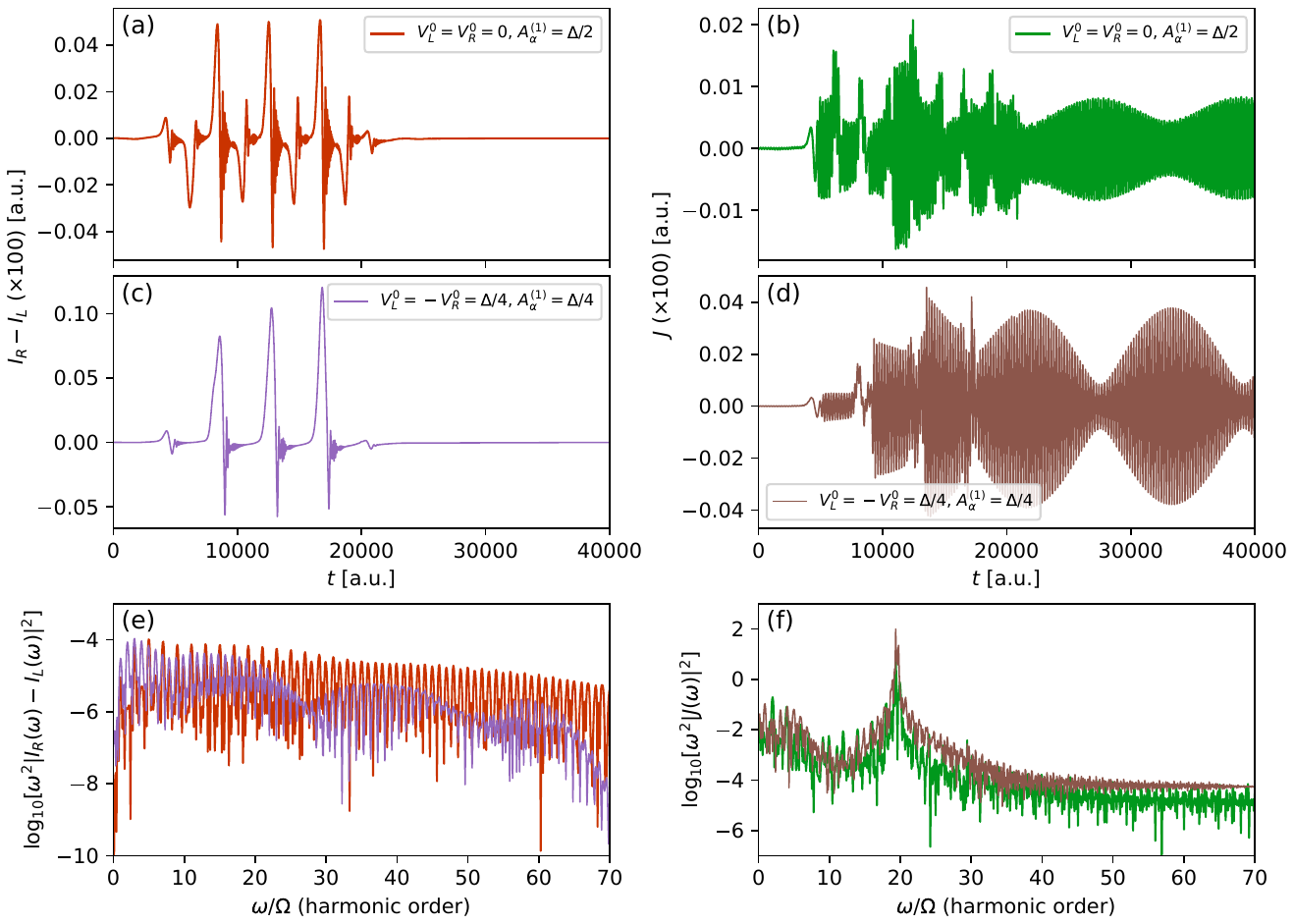}
  \caption{Same as Fig.~\ref{fig:hhg-currents-vis} but with weak coupling to cavity photons of energy $\omega_{\boldsymbol{\mu}}=(0.03,0.03,0.045)$~a.u., cf.~Fig.~5(c) in the main text.}
  \label{fig:hhg-currents-ir}
\end{figure}

In Figs.~\ref{fig:hhg-currents-nopt}, \ref{fig:hhg-currents-vis}, and~\ref{fig:hhg-currents-ir} we show, for comparison, the time-dependent electric current response and the photon flux when the molecular junction is driven by the monoharmonic bias profile according to Fig.~5 in the main text. We observe that without cavity photons the selection between even and odd harmonics is very clearly demonstrated in the electric current signal [see Fig.~\ref{fig:hhg-currents-nopt}(c)]. When coupling to cavity photons, the electric-current responses still possess high-harmonic generation [Figs.~\ref{fig:hhg-currents-vis}(e) and~\ref{fig:hhg-currents-ir}(e)]. The photon-flux response contains the associated harmonics of the basic driving frequency, but the time-resolved signal is heavily masked by higher-frequency oscillations [see Figs.~\ref{fig:hhg-currents-vis}(b,d) and~\ref{fig:hhg-currents-ir}(b,d)]. These higher-frequency oscillations precisely match with the cavity photon energies [see Figs.~\ref{fig:hhg-currents-vis}(f) and~\ref{fig:hhg-currents-ir}(f)], which is also demonstrated by the dipole power spectrum in Fig.~5 of the main text. It is interesting, however, that such peaks are absent in the electric-current response but very clearly appear in the photon-flux response. This is because the electron dynamics at the electrodes is not coupled to the cavity fields, in contrast to the electronic dipole and the photon flux at the molecular region.

\providecommand{\latin}[1]{#1}
\makeatletter
\providecommand{\doi}
  {\begingroup\let\do\@makeother\dospecials
  \catcode`\{=1 \catcode`\}=2 \doi@aux}
\providecommand{\doi@aux}[1]{\endgroup\texttt{#1}}
\makeatother
\providecommand*\mcitethebibliography{\thebibliography}
\csname @ifundefined\endcsname{endmcitethebibliography}
  {\let\endmcitethebibliography\endthebibliography}{}

\end{document}